\documentclass[aps,prd,reprint,superscriptaddress,showpacs,showkeys]{revtex4-1}

\usepackage{amsmath}
\usepackage{amssymb}
\usepackage{natbib}
\usepackage{setspace}

\def\dslash#1{\setbox0=\hbox{$#1$}#1\hskip-\wd0\hbox to\wd0{\hss\sl/\/\hss}}
\makeatletter
\newcommand{\fmslash}[2][0mu]{%
  \mathchoice
    {\fmsl@sh\displaystyle{#1}{#2}}%
    {\fmsl@sh\textstyle{#1}{#2}}%
    {\fmsl@sh\scriptstyle{#1}{#2}}%
    {\fmsl@sh\scriptscriptstyle{#1}{#2}}}
\newcommand{\fmsl@sh}[3]{%
  \m@th\ooalign{$\hfil#1\mkern#2/\hfil$\crcr$#1#3$}}
\makeatother

\begin{document}


\title{Coupling of fermionic fields with mass dimensions one \\ to the O'Raifeartaigh model}

\author{Kai E. Wunderle}
\affiliation{PresiNET Systems, 645 Fort Street, Victoria, BC, V8W 1G2, Canada}
\affiliation{University of Saskatchewan, 116 Science Place, Saskatoon, SK, S7N 5E2, Canada}
\author{Rainer Dick}
\affiliation{University of Saskatchewan, 116 Science Place, Saskatoon, SK, S7N 5E2, Canada}

\date{\today}

\begin{abstract}
The objective of this article is to discuss the coupling of fermionic fields with mass dimension one to the O'Raifeartaigh model to study supersymmetry breaking for fermionic fields with mass dimension one.

We find that the coupled model has two distinct solutions. The first solution represents a local minimum of the superpotential which spontaneously breaks supersymmetry in perfect analogy to the O'Raifeartaigh model. The second solution is more intriguing as it corresponds to a global minimum of the superpotential. In this case the coupling to the fermionic sector restores supersymmetry. However, this is achieved at the cost of breaking Lorentz invariance. Finally, the mass matrices for the multiplets of the coupled model are presented. It turns out that it contains two bosonic triplets and one fermionic doublet which are mass multiplets. In addition it contains a massless fermionic doublet as well as one fermionic triplet which is not a mass multiplet but rather an interaction multiplet that contains component fields of different mass dimension.

These results show that the presented model for fermionic fields with mass dimension one is a viable candidate for supersymmetric dark matter that could be accessible to experiments in the near future.
\end{abstract}


\maketitle


%




\section{\label{Sintro}Introduction}
In a recent publication we present a supersymmetric formalism for fermionic fields with mass dimension one \cite{wunderle10}. It represents a generalization of ELKO spinors that were introduced in Refs. \cite{ahluwaliakhalilova05a} and \cite{ahluwaliakhalilova05b} to the more fundamental concept of fermionic fields with mass dimension one in superspace. In the following article we present a first application of these concepts by constructing a simple toy model. This model conctains the fermionic fields with mass dimension one as supersymmetric extension of the O'Raifeartaigh model.

From the theoretical point of view the study of fermionic fields with mass dimension one proves interesting. However, up to now only few references and calculations involving the general spinor superfield exist in the literature \cite{gates77,siegel79,gates01}. These publications are based on spinor superfields with standard mass dimension and, therefore, do not represent a supersymmetric generalization of ELKO spinors in absence of mass dimension transmuting operators as discussed in \cite{darocha09}.

There are several reasons that make them a good candidate for dark matter (DM). Due to their mass dimension, they only interact very weakly with Standard Model (SM) spinors and gauge fields while the dominant contribution comes from unsuppressed interactions with neutral scalar fields \cite{darocha07}, e.g. the Higgs field in the SM. This is very encouraging as it places them at the edge of detectability of present particle detectors. At the same time this property makes them a first-principle DM candidate \cite{ahluwaliakhalilova05b}. In addition, the model can be mapped to a scalar field theory and it is tempting to consider them as a source of inflation. This could have interesting consequences for cosmological models \cite{boehmer07,boehmer07b,boehmer08,fabbri10}.

Fermionic fields with mass dimension one are also of interest in light of work presented in Ref. \cite{novello07} where it was shown that the spacetime metric can be interpreted as effective geometry without its own dynamics. Instead the metric inherits its dynamics from two fundamental spinor fields.

Other interesting publications on ELKO fields include the investigation of contraints on the spacetime metric and topology caused by the spinor field dynamics \cite{darocha11}, the discussion of ELKO fields in gravity theories with explicit torsional contributions \cite{fabbri12}, as well as the study of ELKO fields in the framework of Weinberg's field theory formalism \cite{gillard10}.

In its modern formulation supersymmetry was introduced by Wess and Zumino \cite{wess74,wess74c,wess74b}. Shortly thereafter, O'Raifeartaigh generalized the Wess-Zumino model by assuming several superfields instead of one superfield \cite{oraifeartaigh75}. With a specific choice of structure constants, he was able to show that his model contains regions in parameter space where at least one of the auxiliary fields acquires a nonvanishing expectation value and thus supersymmetry is spontaneously broken. Since then supersymmetry has been incorporated into more sophisticated models which culminated in the formulation of the Minimally Supersymmetric extension of the Standard Model \cite{drees04,dine07}, superstring theory \citep{green87,green87b,dine07}, and supergravity \cite{wess82}. It has to be emphasized that it is not intended to replace any of these models. Instead, it is used to formulate a supersymmetric extension using fermionic fields with mass dimension one. Due to its properties it can be thought of as a hidden sector extension.

The rest of this publication is structured as follows. First, a brief summary of the most important concepts and results from Ref. \cite{wunderle10} that are necessary for the discussion in this publication are given in Sec. \ref{SVgeneral}. Afterwards in Sec. \ref{SLint}, the interaction Lagrangian that describes the coupling of the fermionic fields with mass dimension one to the O'Raifeartaigh model is derived. Then in Sec. \ref{Scouplingmodelnz}, the coupling of the fermionic field with mass dimension one to the chiral superfield with non-vanishing expectation value is discussed. Finally, the results are  summarized in Sec. \ref{Sconclusion}.
 
\section{\label{SVgeneral}The General Spinor Superfield}
A discussion of the chiral and anti-chiral superfields for fermionic fields with mass dimension one can be found in \cite{wunderle10}. 
 
In analogy to the general scalar superfield, the general spinor superfield can be expanded as expansion in Grassmann variables
\begin{align}
V_\alpha
	&= \kappa_\alpha
	+ \theta^\beta M_{\beta \alpha}
	- \bar{\theta}^{\dot{\beta}} N_{\dot{\beta}\alpha}
	+ \theta^2 \psi_\alpha
	+ \bar{\theta}^2 \chi_\alpha \notag \\
	&\quad + \theta \sigma^\mu \bar{\theta} \omega_{\mu \alpha}
	- \theta^2 \bar{\theta}^{\dot{\beta}} R_{\dot{\beta} \alpha}
	+ \bar{\theta}^2 \theta^\beta S_{\beta \alpha}
	+ \theta^2 \bar{\theta}^2 \lambda_\alpha \, .
\label{Valpha}
\end{align}
Here, $\kappa$, $\psi$, $\chi$, and $\lambda$ are Majorana spinors, $M$, $N$, $R$, and $S$ are complex second-rank spinors, and the spinor vector $\omega$ is equivalent to a complex third-rank spinor. 

The chiral spinor field $X_\alpha$ and anti-chiral spinor field $Y_\alpha$ are then found by repeated operation of the covariant derivative onto the general spinor superfield
\begin{align}
X_\alpha
	&= - \frac{1}{4} \bar{D}^2 V_\alpha \, , \displaybreak[3] \\
Y_\alpha
	&= - \frac{1}{4} D^2 V_\alpha \, .
\end{align}

The Lagrangian can then be written in a very compact form
\begin{align}
\mathcal{L}
	&= \left( X Y + Y X \right)_D + \frac{m}{2} \left( X X + Y Y \right)_F + h. c. \, .
\label{Lcompact}
\end{align}

It can be shown that the component fields $\kappa$, $\lambda$, $M$, $N$, $R$, and $S$ in Eq. (\ref{Valpha}) are not independent. Furthermore, the spinor-vector field $\omega^\mu_\alpha$ is always contracted with a four derivative. After introducing a convenient set of component fields the chiral and anti-chiral superfields can be written as 
\begin{align}
X_\alpha
	&= \chi_\alpha
	+ \theta^\beta \tilde{S}_{\beta \alpha}
	+ \theta^2 \left( \tilde{\lambda}_\alpha + \frac{i}{2} \tilde{\omega}_\alpha \right) \notag \\
	&\quad - i \theta \dslash{\partial} \bar{\theta} \chi_\alpha
	+ \frac{i}{2} \theta^2 \bar{\theta}^{\dot{\gamma}} \bar{\dslash{\partial}}_{\dot{\gamma}}{}^\beta \tilde{S}_{\beta \alpha}
	- \frac{1}{4} \theta^2 \bar{\theta}^2 \Box \chi_\alpha \, , \label{Xchiral}\\
Y_\alpha
	&= \psi_\alpha
	- \bar{\theta}^{\dot{\beta}} \tilde{R}_{\dot{\beta} \alpha}
	+ \bar{\theta}^2 \left( \tilde{\lambda}_\alpha - \frac{i}{2} \tilde{\omega}_\alpha \right)
	+ i \theta \dslash{\partial} \bar{\theta} \psi_\alpha \notag \\
	&\quad + \frac{i}{2} \theta^\gamma \bar{\theta}^2 \dslash{\partial}_\gamma{}^{\dot{\beta}} \tilde{R}_{\dot{\beta} \alpha}
	- \frac{1}{4} \theta^2 \bar{\theta}^2 \Box \psi_\alpha \, , \label{Yantichiral}
\end{align}
and the on-shell Lagrangian is found to be
\begin{align}
\mathcal{L}
	&= \partial_\mu \chi \partial^\mu \psi
	+ \partial_\mu \psi \partial^\mu \chi
	- \frac{m^2}{4} \psi \chi
	- \frac{m^2}{4} \chi \psi \notag \\
	&\quad + \frac{i}{2} \mathrm{Tr}{\left( \tilde{S}^T \dslash{\partial} \tilde{R} \right)}
	+ \frac{i}{2} \mathrm{Tr}{\left( \tilde{R}^T \bar{\dslash{\partial}} \tilde{S} \right)}
	- \frac{m}{4} \mathrm{Tr}{\left( \tilde{S}^T \tilde{S} \right)} \notag \\
	&\quad - \frac{m}{4} \mathrm{Tr}{\left( \tilde{R}^T \tilde{R} \right)} + h. c. \, .
\label{Lonshell}
\end{align}
Here, the bosonic second-rank spinor fields satisfy a Weyl type equation which reconciles the number of bosonic and fermionic on-shell degrees of freedom.

\section{\label{SLint}The Interaction Lagrangian}
\subsection{\label{SSDimAnalysis}Dimensional Analysis}
The building blocks of the Wess-Zumino model and therefore also of the O'Raifeartaigh model have mass dimension
\begin{align}
\mathrm{dim} \, V
	&= 0 \, , \quad 
\mathrm{dim} \, \Phi
	= 1 \, , \quad 
\mathrm{dim} \, D_\alpha
	= \frac{1}{2} \, .
\end{align}
If $\chi$ is identified with a fermionic field with mass dimension one it can be shown that the corresponding spinor superfields and covariant derivatives satisfy
\begin{align}
\mathrm{dim} \, V_\alpha
	&= 0 \, , \quad 
\mathrm{dim} \, X_\alpha
	= 1\, .
\end{align}
As the mass dimension of the general spinor superfield is lower than the one of the general scalar superfield there are more possible contributions to the Lagrangian. This indicates a richer structure. With these results for the mass dimension of the building blocks of the Lagrangian all possible terms can be worked out. For notational simplicity the hermitian conjugate contributions to the Lagrangian were omitted for the following discussion but have to be considered as well.

For the construction of a supersymmetric Lagrangian the contributions have to satisfy three conditions. First, there can be no uncontracted spinor indices. Second, the mass dimension of the structure constants must be positive. Third, the mass dimension of the terms must be appropriate for contributions via $D$- or $F$-component.

The objective is to construct a coupling of the fermionic fields with mass dimension one to the O'Raifeartaigh model. Therefore, contributions containing three superfields have to be considered. This results in two possible scenarios, $\sim \Phi \Phi V_\alpha$ and $\sim \Phi V^\alpha V_\alpha$.

In the first case which is presented in Table \ref{TPhiPhiVInteraction} the number of products is rather short. The two chiral superfields of the O'Raifeartaigh model each have mass dimension one, while the simplest contribution involving the general superfield $V_\alpha$ that has no uncontracted spinor indices is $D V$ with a mass dimension of $1/2$. As $D V$ is not chiral the F-component cannot be used and the mass dimension of $5/2$ for $\Phi \Phi D V$ is too big for a contribution via the D-component. Any other product that can be conceived using two $\Phi$ and one $V_\alpha$, as well as an appropriate number of covariant derivatives yields a mass dimension in excess of 3.

\begin{table}
\caption{\label{TPhiPhiVInteraction} Contributions to the Lagrangian for a coupling between two chiral superfields of the O'Raifeartaigh model $\Phi$ and one general superfield with one free spinor index. In addition to the contributions built from products of unbarred superfields, the hermitian conjugates are permitted as well.}
\begin{ruledtabular}
\begin{tabular}{lll}
Product & $\mathrm{dim}_M$ & Contributions \\
\hline
$\Phi \Phi D V$ & $5/2$ & not D-component
\end{tabular}
\end{ruledtabular}
\end{table}

Dimensional analysis for the second case reveals that there are numerous possibilities for a coupling involving the D- and F-component. They can be categorized into three distinct groups based on the mass dimension of the superfield products without structure constants. It has to be emphasized that the contributions presented in Table \ref{TPhiVVInteraction} only discusses terms that can be constructed utilizing the chiral superfield $\Phi$, the kinetic superfield $\mathrm{T} \Phi$, and the general superfield $V_\alpha$, as well as its covariant derivatives while terms containing linear derivatives of $\Phi$ were ignored. This restriction was made to preserve the fact that the O'Raifeartaigh model is built solely using the chiral superfield $\Phi$ as well as the kinetic superfield $\mathrm{T} \Phi$.

\begin{table}
\caption{\label{TPhiVVInteraction}Contributions to the Lagrangian for a coupling between one chiral superfields of the O'Raifeartaigh model $\Phi$ and two general superfield with one free spinor index. In addition to the contributions built from products of unbarred superfields, the hermitian conjugates are permitted as well.}
\begin{ruledtabular}
\begin{tabular}{lll}
Product & $\mathrm{dim}_M$ & Contributions \\
\hline
$\Phi V V$ & 1 & $\left( m \Phi V V \right)_D$, $\left( m \bar{\Phi} V V \right)_D$ \\
$\mathrm{T} \Phi V V$ & 2 & $\left( \mathrm{T} \Phi V V \right)_D$,$\left( \mathrm{T} \bar{\Phi} V V \right)_D$ \\
$\Phi X V$ & 2 & $\left( \Phi X V \right)_D$,$\left( \bar{\Phi} X V \right)_D$,$\left( \Phi Y V \right)_D$,$\left( \bar{\Phi} Y V \right)_D$ \\
$\Phi D V \! D V$ & 2 & $\left( \Phi D V D V \right)_D$, $\left( \bar{\Phi} D V D V \right)_D$  \\
$\Phi V X$ & 2 & $\left( \Phi V X \right)_D$,$\left( \bar{\Phi} V X \right)_D$,$\left( \Phi V Y \right)_D$,$\left( \bar{\Phi} V Y \right)_D$ \\
$\Phi X X$ & 3 & $\left( \Phi X X \right)_F$,$\left( \bar{\Phi} Y Y \right)_F$
\end{tabular}
\end{ruledtabular}
\end{table}

The first group summarizing all terms with mass dimension one contains only two terms as well as their hermitian conjugates. The first term of this group is the product of the chiral superfield $\Phi$ and two general superfields $V_\alpha$, while the second one is the product of the anti-chiral superfield $\bar{\Phi}$ and two general superfields $V_\alpha$. As $V_\alpha$ is neither chiral nor anti-chiral only contributions via the D-component are possible.

The second group collects all terms with mass dimension two containing two covariant derivatives in addition to the field configuration of the first group. Within the outlined framework it is possible to construct 12 distinct terms which are all contributing via the D-component as neither $V_\alpha$ nor $D V$ are chiral or anti-chiral.

Of special interest is the third group summarizing all terms with mass dimension 3. Due to its mass dimension it can only contain contributions via the $F$-component. Indeed, it is possible to construct two such contributions. They are very similar to the mass terms for the fermionic fields with mass dimension one $\left( m X X \right)_F$ and $\left( m Y Y \right)_F$ which potentially leads to a connection between the vacuum expectation value of the spontaneously broken superfield in the O'Raifeartaigh model and the mass of the fermionic fields with mass dimension one.

\subsection{\label{SScouplingmodel}The Lagrangian}
Even though there is a large number of potential contributions to the Lagrangian, the following discussion will be restricted to the most promising ones which are those contributing via the $F$-component. This reduces the number of terms from 32 to 4 if the hermitian conjugates are considered as well. Of the remaining terms two contain only chiral superfields, $\Phi X X$ and $\Phi \bar{Y} \bar{Y}$, while $\bar{\Phi} \bar{X} \bar{X}$ and $\bar{\Phi} Y Y$ are a product of anti-chiral superfields. At the present point the index that distinguishes the three distinct chiral superfields $\Phi_j$ of the O'Raifeartaigh model is treated generally. Later on it will be restricted to describe coupling of the chiral superfields $X_\alpha$ and the anti-chiral superfields $Y_\alpha$ to a specific superfield of the O'Raifeartaigh model, e.g. $\Phi_3$.

In terms of superfield products the Lagrangian can be expressed in a very compact form
\begin{align}
\mathcal{L}
	&= \frac{1}{2} \left( \Phi_a \cdot \mathrm{T} \Phi_a \right)_F 
	- \lambda_a \left( \Phi_a \right)_F
	- \frac{1}{2} M_{a b} \left( \Phi_a \cdot \Phi_b \right)_F \notag \\
	&\quad - \frac{1}{3} g_{a b c} \left( \Phi_a \cdot \Phi_b \cdot \Phi_c \right)_F
	+ \left( X^\alpha Y_\alpha \right)_D 
	+ \left( Y^\alpha X_\alpha \right)_D \notag \\
	&\quad + \frac{m}{2} \left( X^\alpha X_\alpha \right)_F
	+ \frac{m}{2} \left( Y^\alpha Y_\alpha \right)_F 
	+ \xi \left( \Phi_j X^\alpha X_\alpha \right)_F \notag \\
	&\quad + \xi \left( \bar{\Phi}_j Y^\alpha Y_\alpha \right)_F
	+ h. c. \, ,
\end{align}
where the strength of the interaction is encoded in the real coupling constant $\xi$.
Alternatively, the Lagrangian can be written in terms of the superfield components
\begin{widetext}
\begin{align}
\mathcal{L}
	&= - \frac{1}{2} A_a \Box A^\dagger_a 
	+ \frac{1}{2} F_a F^\dagger_a 
	+ \frac{i}{4} \phi_a \dslash{\partial} \bar{\phi}_a 
	- \lambda_a F_a 
	- \frac{1}{2} M_{a b} \left( A_a F_b + F_a A_b - \frac{1}{2} \phi_a \phi_b \right) \notag \\
	&\quad - \frac{1}{3} g_{a b c} \left( A_a A_b F_c 
	+ A_a F_b A_c 
	+ F_a A_b A_c
	- \frac{3}{2} A_a \phi_b \phi_c \right)
	- \frac{1}{2} \Box A_a A^\dagger_a 
	+ \frac{1}{2} F_a F^\dagger_a 
	+ \frac{i}{4} \bar{\phi}_a \bar{\dslash{\partial}} \phi_a
	- \lambda_a F^\dagger_a \notag \displaybreak[3] \\
	&\quad - \frac{1}{2} M_{a b} \left( A^\dagger_a F^\dagger_b + F^\dagger_a A^\dagger_b - \frac{1}{2} \bar{\phi}_a \bar{\phi}_b \right)
	- \frac{1}{3} g_{a b c} \left( A^\dagger_a A^\dagger_b F^\dagger_c 
	+ A^\dagger_a F^\dagger_b A^\dagger_c 
	+ F^\dagger_a A^\dagger_b A^\dagger_c
	- \frac{3}{2} A^\dagger_a \bar{\phi}_b \bar{\phi}_c \right)
	+ 2 \partial_\mu \chi \partial^\mu \psi
	+ 2 \tilde{\lambda} \tilde{\lambda} \notag \displaybreak[3] \\
	&\quad + \frac{1}{2} \tilde{\omega} \tilde{\omega}
	+ m \chi \tilde{\lambda}
	+ \frac{i m}{2} \chi \tilde{\omega}
	+ m \psi \tilde{\lambda}
	- \frac{i m}{2} \psi \tilde{\omega} 
	+ \frac{i}{2} \mathrm{Tr}{\left( \tilde{S}^T \dslash{\partial} \tilde{R} \right)} 
	+ \frac{i}{2} \mathrm{Tr}{\left( \tilde{R}^T \bar{\dslash{\partial}} \tilde{S} \right)}
	- \frac{m}{4} \mathrm{Tr}{\left( \tilde{S}^T \tilde{S} \right)}
	- \frac{m}{4} \mathrm{Tr}{\left( \tilde{R}^T \tilde{R} \right)} \notag \displaybreak[3] \\
	&\quad + 2 \partial_\mu \bar{\chi} \partial^\mu \bar{\psi}
	+ 2 \tilde{\bar{\lambda}} \tilde{\bar{\lambda}}
	+ \frac{1}{2} \tilde{\bar{\omega}} \tilde{\bar{\omega}}
	+ m \bar{\chi} \tilde{\bar{\lambda}}
	- \frac{i m}{2} \bar{\chi} \tilde{\bar{\omega}}
	+ m \bar{\psi} \tilde{\bar{\lambda}}
	+ \frac{i m}{2} \bar{\psi} \tilde{\bar{\omega}}
	+ \frac{i}{2} \mathrm{Tr}{\left( \tilde{\bar{S}}^T \bar{\dslash{\partial}} \tilde{\bar{R}} \right)}
	+ \frac{i}{2} \mathrm{Tr}{\left( \tilde{\bar{R}}^T \dslash{\partial} \tilde{\bar{S}} \right)}
	- \frac{m}{4} \mathrm{Tr}{\left( \tilde{\bar{S}}^T \tilde{\bar{S}} \right)} \notag \displaybreak[3] \\
	&\quad - \frac{m}{4} \mathrm{Tr}{\left( \tilde{\bar{R}}^T \tilde{\bar{R}} \right)} 
	+ 2 \xi A_j \chi \lambda
	+ i \xi A_j \chi \tilde{\omega}
	- \frac{\xi}{2} A_j \mathrm{Tr}{\left( \tilde{S}^T \tilde{S} \right)}
	- \xi \phi_j \tilde{S} \chi
	+ \xi F_j \chi \chi
	+ 2 \xi A_j^\dagger \bar{\chi} \bar{\lambda}
	- i \xi A_j^\dagger \bar{\chi} \tilde{\bar{\omega}}
	- \frac{\xi}{2} A_j^\dagger \mathrm{Tr}{\left( \tilde{\bar{S}}^T \tilde{\bar{S}} \right)} \notag \displaybreak[3] \\
	&\quad + \xi \bar{\phi}_j \tilde{\bar{S}} \bar{\chi}
	+ \xi F_j^\dagger \bar{\chi} \bar{\chi}
	+ 2 \xi A_j^\dagger \psi \lambda
	- i \xi A_j^\dagger \psi \tilde{\omega}
	- \frac{\xi}{2} A_j^\dagger \mathrm{Tr}{\left( \tilde{R}^T \tilde{R} \right)}
	- \xi \bar{\phi}_j \tilde{R} \psi
	+ \xi F_j^\dagger \psi \psi
	+ 2 \xi A_j \bar{\psi} \bar{\lambda}
	+ i \xi A_j \bar{\psi} \tilde{\bar{\omega}} \notag \\
	&\quad - \frac{\xi}{2} A_j \mathrm{Tr}{\left( \tilde{\bar{R}}^T \tilde{\bar{R}} \right)}
	+ \xi \phi_j \tilde{\bar{R}} \bar{\psi}
	+ \xi F_j \bar{\psi} \bar{\psi} \, .
\label{CouplingL}
\end{align}
\end{widetext}
\subsection{The Equations of Motion for the Auxiliary Fields}
A look at the Lagrangian in Eq. (\ref{CouplingL}) reveals that the fields $F_a$, $\tilde{\lambda}$, and $\tilde{\omega}$ are auxiliary fields and thus can be eliminated from the Lagrangian using their respective equations of motion. In addition to the equations of motion for the auxiliary fields the equations of motion for the spinor fields $\phi_j$ and the second rank spinor fields $\tilde{R}$ and $\tilde{S}$ were derived as well
\begin{align}
&F_d
	= \lambda_d
	+ M_{d a} A^\dagger_a
	+ g_{d a b} A^\dagger_a A^\dagger_b
	- \xi \delta_{d j} \left( \bar{\chi} \bar{\chi} + \psi \psi \right) \, , \label{eqmF} \\
&\tilde{\lambda}_\alpha
	= - \frac{m}{4} \left( \chi_\alpha + \psi_\alpha \right)
	- \xi \frac{1}{2} \left( A_j \chi_\alpha + A_j^\dagger \psi_\alpha \right) \, , \label{eqmlambda} \displaybreak[3] \\
&\tilde{\omega}_\alpha
	= - \frac{i m}{2} \left( \chi_\alpha - \psi_\alpha \right)
	- i \xi \left( A_j \chi_\alpha - A_j^\dagger \psi_\alpha \right) \, , \label{eqmomega} \displaybreak[3] \\
&\frac{i}{2} \bar{\dslash{\partial}}_{\dot{\alpha} \beta} \phi^\beta_d
	= \frac{1}{2} M_{d a} \bar{\phi}_{\dot{\alpha} a}
	+ g_{d a b} A^\dagger_a \bar{\phi}_{\dot{\alpha} b} \notag \\
	&\qquad\qquad + \xi \delta_{d j} \left( \tilde{\bar{S}}_{\dot{\alpha} \dot{\gamma}} \bar{\chi}^{\dot{\gamma}} + \tilde{R}_{\dot{\alpha} \gamma} \psi^\gamma \right) \, , \label{eqmphi} \displaybreak[3] \\
&\left( \frac{m}{2} + \xi A_j^\dagger \right) \tilde{R}_{\dot{\beta} \alpha}
	= - i \bar{\dslash{\partial}}_{\dot{\beta} \gamma} \tilde{S}^\gamma{}_\alpha
	- \xi \bar{\phi}_{j \dot{\beta}} \psi_\alpha \, , \label{eqmR} \\
&\left( \frac{m}{2} + \xi A_j \right) \tilde{S}_{\beta \alpha}
	= i \dslash{\partial}_{\beta \dot{\delta}} \tilde{R}^{\dot{\delta}}{}_\alpha
	- \xi \phi_{j \beta} \chi_\alpha \, . \label{eqmS}
\end{align}

\subsection{Spontaneous Symmetry Breaking}
To determine whether or not supersymmetry is spontaneously broken the Lagrangian describing the coupling between the O'Raifeartaigh model and the fermionic sector has to be split up into the super kinetic term $\mathcal{L}_\text{kin}$ and the superpotential $\mathcal{L}_\text{pot}$. The super-kinetic term is given by
\begin{align}
\mathcal{L}_\text{kin}
	&= - \frac{1}{2} A_a \Box A^\dagger_a 
	- \frac{1}{2} \Box A_a A^\dagger_a 
	+ \frac{i}{4} \phi_a \dslash{\partial} \bar{\phi}_a
	+ \frac{i}{4} \bar{\phi}_a \bar{\dslash{\partial}} \phi_a \notag \\
	&\quad + 2 \partial_\mu \chi \partial^\mu \psi
	+ 2 \partial_\mu \bar{\chi} \partial^\mu \bar{\psi} 
	+ \frac{i}{2} \mathrm{Tr}{\left( \tilde{S}^T \dslash{\partial} \tilde{R} \right)}\notag \\
	&\quad + \frac{i}{2} \mathrm{Tr}{\left( \tilde{R}^T \bar{\dslash{\partial}} \tilde{S} \right)}
	+ \frac{i}{2} \mathrm{Tr}{\left( \tilde{\bar{S}}^T \bar{\dslash{\partial}} \tilde{\bar{R}} \right)}
	+ \frac{i}{2} \mathrm{Tr}{\left( \tilde{\bar{R}}^T \dslash{\partial} \tilde{\bar{S}} \right)} \, .
\label{CouplingLkin}
\end{align}
The remaining terms of the Lagrangian make up the superpotential. However, it still needs to be expressed solely in terms of the auxiliary fields $F_a$, $\tilde{\lambda}$, and $\tilde{\omega}$.

It turns out that the combination of component fields in the prefactors of $\tilde{\omega}$ and $\tilde{\lambda}$ correspond to the equations of motion of $\tilde{\omega}$ and $\tilde{\lambda}$ which simplifies the superpotential significantly. In addition, collecting all terms proportional to $F_a$ and $F_a^\dagger$ reveals that the combinations of component fields making up the prefactors correspond to $F^\dagger_a$ and $F_a$ respectively.

Up to now only the equations of motion for the auxiliary fields were used to rewrite the superpotential in terms of the auxiliary fields. However, the intermediate result for the superpotential still depends on $A_j$, $\phi_j$, $\tilde{S}$, and $\tilde{R}$.  At this point the equations of motion for these fields have to be used to eliminate or rewrite the terms of interest. Using the equation of motion for $\phi_a$ and $\bar{\phi}_a$ the terms containing $\phi_a$ and its hermitian conjugate are reduced to two kinetic terms. The same can be repeated using the equations of motion for the second rank spinor fields $\tilde{S}$ and $\tilde{R}$ and the resulting superpotential has an particularly simple form
\begin{align}
\mathcal{L}_\text{pot}	
	&= F_a F^\dagger_a
	+ \frac{i}{2} \phi_a \dslash{\partial} \bar{\phi}_a
	+ \frac{i}{2} \bar{\phi}_a \bar{\dslash{\partial}} \phi_a
	+ 2 \tilde{\lambda} \tilde{\lambda}
	+ \frac{1}{2} \tilde{\omega} \tilde{\omega} \notag \\
	&\quad + \frac{i}{2} \mathrm{Tr}{\left( \tilde{R}^T \bar{\dslash{\partial}} \tilde{S} \right)}
	+ \frac{i}{2} \mathrm{Tr}{\left( \tilde{S}^T \dslash{\partial} \tilde{R} \right)} 
	+ 2 \tilde{\bar{\lambda}} \tilde{\bar{\lambda}}
	+ \frac{1}{2} \tilde{\bar{\omega}} \tilde{\bar{\omega}} \notag \\
	&\quad 
	+ \frac{i}{2} \mathrm{Tr}{\left( \tilde{\bar{R}}^T \dslash{\partial} \tilde{\bar{S}} \right)}
	+ \frac{i}{2} \mathrm{Tr}{\left( \tilde{\bar{S}}^T \bar{\dslash{\partial}} \tilde{\bar{R}} \right)} \, .
\end{align}
It can be seen that the superpotential corresponds only approximately to the actual potential and still contains numerous kinetic terms for the bosonic component fields $\tilde{R}$ and $\tilde{S}$ as well as the fermionic component fields $\phi_a$. As the super-kinetic term in Eq. (\ref{CouplingLkin}) is by construction free of contributions to the superpotential the actual superpotential $U$ is found to be
\begin{align}
U
	&= F_a F^\dagger_a 
	+ 2 \left( \tilde{\lambda} \tilde{\lambda}
	+ \tilde{\bar{\lambda}} \tilde{\bar{\lambda}} \right)
	+ \frac{1}{2} \left( \tilde{\omega} \tilde{\omega}
	+ \tilde{\bar{\omega}} \tilde{\bar{\omega}} \right) \, .
\end{align}
This superpotential is in perfect analogy to the superpotential for the O'Raifeartaigh model in Ref. \cite{oraifeartaigh75}. Besides the term induced by the bosonic auxiliary fields $F_a$ it includes two additional terms for the auxiliary fields $\tilde{\lambda}$ and $\tilde{\omega}$ which originate in the model for fermionic fields with mass dimension one that were used to extend the O'Raifeartaigh model. As the contributions from $\tilde{\lambda}$ and $\tilde{\omega}$ are given by a sum of spinor products it is not immediately clear whether the superpotential is always positive. The sum of spinor products and their hermitian conjugates is real, however, it is not sufficient to conclude that they are positive as well. Nevertheless, there are two arguments that should guarantee a positive superpotential. First, the construction of the fermionic sector using the supersymmetry algebra ensures a positive energy spectrum. For the O'Raifeartaigh model this property is well established and the positivity of the energy spectrum for the fermionic fields with mass dimension one was shown in Ref. \cite{wunderle10}. Second, the coupling of two theories with positive energy spectrum should possess the same fundamental property. Therefore, if the expectation values for all auxiliary fields $F_a$, $\tilde{\lambda}$, and $\tilde{\omega}$ vanish supersymmetry is preserved. Otherwise the superpotential acquires a finite positive minimum and supersymmetry is spontaneously broken.

\subsection{The On-shell Lagrangian}
The calculation of the on-shell Lagrangian is very similar to the discussion in the previous section. However, this time the equations of motion for the auxiliary fields are used to eliminate the auxiliary fields from the superpotential. Inserting Eqs. (\ref{eqmF}) - (\ref{eqmomega}) into Eq. (\ref{CouplingL}) leads to
\begin{widetext}
\begin{align}
\mathcal{L}
	&= - \frac{1}{2} A_a \Box A^\dagger_a 
	- \frac{1}{2} \Box A_a A^\dagger_a 
	+ \frac{i}{4} \phi_a \dslash{\partial} \bar{\phi} 
	+ \frac{i}{4} \bar{\phi}_a \bar{\dslash{\partial}} \phi_a
	+ \partial_\mu \chi \partial^\mu \psi
	+ \partial_\mu \psi \partial^\mu \chi
	+ \partial_\mu \bar{\chi} \partial^\mu \bar{\psi}
	+ \partial_\mu \bar{\psi} \partial^\mu \bar{\chi}
	+ \frac{i}{2} \mathrm{Tr}{\left( \tilde{S}^T \dslash{\partial} \tilde{R} \right)}\notag \\
	&\quad + \frac{i}{2} \mathrm{Tr}{\left( \tilde{R}^T \bar{\dslash{\partial}} \tilde{S} \right)}
	+ \frac{i}{2} \mathrm{Tr}{\left( \tilde{\bar{S}}^T \bar{\dslash{\partial}} \tilde{\bar{R}} \right)}
	+ \frac{i}{2} \mathrm{Tr}{\left( \tilde{\bar{R}}^T \dslash{\partial} \tilde{\bar{S}} \right)}
	- \lambda_a \lambda_a
	- \lambda_a M_{a d} A_d
	- \lambda_a g_{a d e} A_d A_e
	+ \xi \lambda_a \delta_{a j} \left( \chi \chi + \bar{\psi} \bar{\psi} \right) \notag \\
	&\quad - \lambda_a M_{a b} A^\dagger_b
	- M_{a b} M_{a d} A^\dagger_b A_d
	- M_{a b} g_{a d e} A^\dagger_b A_d A_e
	+ \xi M_{a b} A^\dagger_b \delta_{a j} \left( \chi \chi + \bar{\psi} \bar{\psi} \right) 
	- \lambda_a g_{a b c} A^\dagger_b A^\dagger_c
	- M_{a d} g_{a b c} A^\dagger_b A^\dagger_c A_d \notag \\
	&\quad - g_{a b c} g_{a d e} A^\dagger_b A^\dagger_c A_d A_e
	+ \xi g_{a b c} A^\dagger_b A^\dagger_c \delta_{a j} \left( \chi \chi + \bar{\psi} \bar{\psi} \right)
	+ \xi \lambda_a \delta_{a j} \left( \bar{\chi} \bar{\chi} + \psi \psi \right)
	+ \xi M_{a d} \delta_{a j} \left( \bar{\chi} \bar{\chi} + \psi \psi \right) A_d \notag \\
	&\quad + \xi g_{a d e} \delta_{a j} \left( \bar{\chi} \bar{\chi} + \psi \psi \right) A_d A_e		- \xi^2 \delta_{a j} \delta_{a j} \left( \bar{\chi} \bar{\chi} + \psi \psi \right) \left( \chi \chi + \bar{\psi} \bar{\psi} \right)
	+ \frac{1}{4} M_{a b} \phi_a \phi_b
	+ \frac{1}{4} M_{a b} \bar{\phi}_a \bar{\phi}_b  
	+ \frac{1}{2} g_{a b c} A_a \phi_b \phi_c \notag \\
	&\quad + \frac{1}{2} g_{a b c} A^\dagger_a \bar{\phi}_b \bar{\phi}_c
	- 2 \left( \frac{m}{2} + \xi A_j \right) \left( \frac{m}{2} + \xi A_j^\dagger \right) \chi \psi
	- \left( \frac{m}{4} + \frac{\xi}{2} A_j \right) \mathrm{Tr}{\left( \tilde{S}^T \tilde{S} \right)}
	- \left( \frac{m}{4} + \frac{\xi}{2} A_j^\dagger \right) \mathrm{Tr}{\left( \tilde{R}^T \tilde{R} \right)} \notag \\
	&\quad - 2 \left( \frac{m}{2} + \xi A_j \right) \left( \frac{m}{2} + \xi A_j^\dagger \right) \bar{\chi} \bar{\psi}
	- \left( \frac{m}{4} + \frac{\xi}{2} A_j^\dagger \right) \mathrm{Tr}{\left( \tilde{\bar{S}}^T \tilde{\bar{S}} \right)}
	- \left( \frac{m}{4} + \frac{\xi}{2} A_j \right) \mathrm{Tr}{\left( \tilde{\bar{R}}^T \tilde{\bar{R}} \right)}
	- \xi \phi_j \tilde{S} \chi
	+ \xi \bar{\phi}_j \tilde{\bar{S}} \bar{\chi} \notag \\
	&\quad - \xi \bar{\phi}_j \tilde{R} \psi
	+ \xi \phi_j \tilde{\bar{R}} \bar{\psi} \, .
\label{CouplingLonshell}
\end{align}
\end{widetext}
This is the most general Lagrangian describing the coupling of the O'Raifeartaigh model to a model for fermionic fields with mass dimension one. Up to now no assumptions besides the usual symmetry properties were made regarding the structure constants of the O'Raifeartaigh model. Furthermore, the coupling of the fermionic sector to the O'Raifeartaigh is not restricted to a specific superfield.

\section{\label{Scouplingmodelnz}Coupling to the Field with Nonzero Expectation Value}
The O'Raifeartaigh model contains three chiral superfields. For the specific choice of coupling constants 
\begin{align}
\lambda_3 &= \Lambda \, \text{, all other} \, \Lambda_a = 0 \, , \label{ORaifeartaighconstlambda} \\
m_{1 2}
	&= m_{2 1} = M \, \text{, all other} \, m_{a b} = 0 \, , \label{ORaifeartaighconstM} \\
g_{1 1 3}
	&= g_{1 3 1} = g_{3 1 1} = g \, \text{, all other} \, g_{a b c} = 0 \, , \label{ORaifeartaighconstg}
\end{align}
where $\Lambda$, $M$, and $g$ are real, only one obtains a nonvanishing expectation value. Therefore, there are two possibilities to couple the fermionic sector to the O'Raifeartaigh model. It can either be coupled to the superfield with nonvanishing expectation value -- in this case $\Phi_3$ -- or to one of the superfields with vanishing expectation value. Without loss of generality it is sufficient to discuss one of the possible scenarios as the results for the other case can be obtained in perfect analogy. It can be shown that differences between the scenarios are restricted to the matrix components of the mass matrices while the fundamental properties, e.g. spontaneous supersymmetry breaking, are preserved. 

For convenience the coupling to the chiral superfield with nonvanishing expectation value is discussed. The structure constants remain those introduced in Eqs. (\ref{ORaifeartaighconstlambda}) to (\ref{ORaifeartaighconstg}). For this specific choice the on-shell Lagrangian from Eq. (\ref{CouplingLonshell}) is given by
\begin{widetext}
\begin{align}
\mathcal{L}
	&= - \frac{1}{2} A_1 \Box A^\dagger_1 
	- \frac{1}{2} A_2 \Box A^\dagger_2 
	- \frac{1}{2} A_3 \Box A^\dagger_3 
	- \frac{1}{2} \Box A_1 A^\dagger_1
	- \frac{1}{2} \Box A_2 A^\dagger_2 
	- \frac{1}{2} \Box A_3 A^\dagger_3 
	+ \frac{i}{4} \phi_1 \dslash{\partial} \bar{\phi}_1
	+ \frac{i}{4} \phi_2 \dslash{\partial} \bar{\phi}_2
	+ \frac{i}{4} \phi_3 \dslash{\partial} \bar{\phi}_3 \notag \\
	&\quad + \frac{i}{4} \bar{\phi}_1 \bar{\dslash{\partial}} \phi_1
	+ \frac{i}{4} \bar{\phi}_2 \bar{\dslash{\partial}} \phi_2
	+ \frac{i}{4} \bar{\phi}_3 \bar{\dslash{\partial}} \phi_3
	+ 2 \partial_\mu \chi \partial^\mu \psi
	+ 2 \partial_\mu \bar{\chi} \partial^\mu \bar{\psi} 
	+ \frac{i}{2} \mathrm{Tr}{\left( \tilde{S}^T \dslash{\partial} \tilde{R} \right)}
	+ \frac{i}{2} \mathrm{Tr}{\left( \tilde{R}^T \bar{\dslash{\partial}} \tilde{S} \right)}
	+ \frac{i}{2} \mathrm{Tr}{\left( \tilde{\bar{S}}^T \bar{\dslash{\partial}} \tilde{\bar{R}} \right)} \notag \\
	&\quad + \frac{i}{2} \mathrm{Tr}{\left( \tilde{\bar{R}}^T \dslash{\partial} \tilde{\bar{S}} \right)} 
	- \Lambda^2
	- \Lambda g A_1 A_1
	+ \xi \Lambda \left( \chi \chi + \bar{\psi} \bar{\psi} \right)
	- M^2 A^\dagger_2 A_2
	- M^2 A^\dagger_1 A_1
	- 2 M g A^\dagger_2 A_1 A_3
	- \Lambda g A^\dagger_1 A^\dagger_1 \notag \\
	&\quad - M g A^\dagger_2 A^\dagger_1 A_3
	- M g A^\dagger_2 A^\dagger_3 A_1
	- 4 g^2 A^\dagger_1 A^\dagger_3 A_1 A_3
	- g^2 A^\dagger_1 A^\dagger_1 A_1 A_1
	+ \xi g A^\dagger_1 A^\dagger_1 \left( \chi \chi + \bar{\psi} \bar{\psi} \right) 
	+ \xi \Lambda \left( \bar{\chi} \bar{\chi} + \psi \psi \right) \notag \\
	&\quad + \xi g \left( \bar{\chi} \bar{\chi} + \psi \psi \right) A_1 A_1
	- \xi^2 \left( \bar{\chi} \bar{\chi} + \psi \psi \right) \left( \chi \chi + \bar{\psi} \bar{\psi} \right)
	+ \frac{1}{2} M \phi_1 \phi_2
	+ \frac{1}{2} M \bar{\phi}_1 \bar{\phi}_2  
	+ g A_1 \phi_1 \phi_3
	+ \frac{1}{2} g A_3 \phi_1 \phi_1 \notag \\
	&\quad + g A^\dagger_1 \bar{\phi}_1 \bar{\phi}_3
	+ \frac{1}{2} g A^\dagger_3 \bar{\phi}_1 \bar{\phi}_1
	- 2 \left( \frac{m}{2} + \xi A_3 \right) \left( \frac{m}{2} + \xi A_3^\dagger \right) \chi \psi
	- \left( \frac{m}{4} + \frac{\xi}{2} A_3 \right) \mathrm{Tr}{\left( \tilde{S}^T \tilde{S} \right)}
	- \left( \frac{m}{4} + \frac{\xi}{2} A_3^\dagger \right) \mathrm{Tr}{\left( \tilde{R}^T \tilde{R} \right)} \notag \\
	&\quad - 2 \left( \frac{m}{2} + \xi A_3 \right) \left( \frac{m}{2} + \xi A_3^\dagger \right) \bar{\chi} \bar{\psi}
	- \left( \frac{m}{4} + \frac{\xi}{2} A_3^\dagger \right) \mathrm{Tr}{\left( \tilde{\bar{S}}^T \tilde{\bar{S}} \right)}
	- \left( \frac{m}{4} + \frac{\xi}{2} A_3 \right) \mathrm{Tr}{\left( \tilde{\bar{R}}^T \tilde{\bar{R}} \right)}
	- \xi \phi_3 \tilde{S} \chi
	+ \xi \bar{\phi}_3 \tilde{\bar{S}} \bar{\chi} \notag \\
	&\quad - \xi \bar{\phi}_3 \tilde{R} \psi
	+ \xi \phi_3 \tilde{\bar{R}} \bar{\psi} \, .
\end{align}
\end{widetext}
The corresponding equations of motion for the auxiliary fields from Eqs. (\ref{eqmF}) to (\ref{eqmomega}) are
\begin{align}
F_1
	&= M A^\dagger_2
	+ g \left( A^\dagger_1 A^\dagger_3 + A^\dagger_3 A^\dagger_1 \right) \, , \label{CouplingeqmF1} \\
F_2
	&= M A^\dagger_1 \, , \label{CouplingeqmF2} \displaybreak[3] \\
F_3
	&= \Lambda
	+ g A^\dagger_1 A^\dagger_1
	- \xi \left( \bar{\chi} \bar{\chi} + \psi \psi \right) \, , \label{CouplingeqmF3} \displaybreak[3] \\
\tilde{\lambda}_\alpha
	&= - \frac{\xi}{2} \left( \frac{m}{2 \xi} + A_3 \right) \chi_\alpha
	- \frac{\xi}{2} \left( \frac{m}{2 \xi} + A_3^\dagger \right) \psi_\alpha \, , \label{Couplingeqmlambda} \\
\tilde{\omega}_\alpha
	&= - i \xi \left( \frac{m}{2 \xi} + A_3 \right) \chi_\alpha
	+ i \xi \left( \frac{m}{2 \xi} \psi_\alpha + A_3^\dagger \right) \psi_\alpha \, . \label{Couplingeqmomega}
\end{align}

\subsection{Limit for Vanishing Interaction}
Before the coupling of the fermionic fields with mass dimension one to the O'Raifeartaigh model is discussed in detail it is important to verify whether the previous results for the Lagrangian and the equations of motion are reasonable. This can be achieved by discussing the special case for a vanishing coupling constant. For $\xi \to 0$ the equations of motion reduce to
\begin{align}
F_1
	&= M A^\dagger_2
	+ g \left( A^\dagger_1 A^\dagger_3 + A^\dagger_3 A^\dagger_1 \right) \, , \\
F_2
	&= M A^\dagger_1 \, , \displaybreak[3] \\
F_3
	&= \Lambda
	+ g A^\dagger_1 A^\dagger_1 \, , \displaybreak[3] \\
\tilde{\lambda}_\alpha
	&= - \frac{m}{4} \left( \chi_\alpha + \psi_\alpha \right) \, , \\
\tilde{\omega}_\alpha
	&= - \frac{i m}{2} \left( \chi_\alpha - \psi_\alpha \right) \, .
\end{align}
In this limit the equations of motion for $F_a$ clearly decouple from those for $\tilde{\lambda}$ and $\tilde{\omega}$. Furthermore, it can be seen that the equations of motion for $F_a$ reproduce the equations of motion of the O'Raifeartaigh model in Ref. \cite {oraifeartaigh75} while the equations of motion for $\tilde{\lambda}$ and $\tilde{\omega}$ are exactly those derived for the model describing fermionic fields with mass dimension one in Ref. \cite{wunderle10}.

As the equations of motion are exactly those of the two individual models the discussion of the expectation values is straightforward. For the bosonic component fields $A_a$ it is found that 
\begin{align}
\left\langle A_1 \right\rangle 
	&= \left\langle A_2 \right\rangle 
	= 0 \, , \, 
	\left\langle A_3 \right\rangle 
	= c_A \, ,
\end{align}
where $c_A$ is a real constant. Therefore, the auxiliary field $F_3$ acquires a nonvanishing expectation value which implies that supersymmetry is spontaneously broken. Furthermore, the fermionic auxiliary fields $\tilde{\lambda}$ and $\tilde{\omega}$ only have a trivial solution
\begin{align}
\left\langle \chi_\alpha \right\rangle 
	&= \left\langle \psi_\alpha \right\rangle 
	= 0 \, .
\end{align}
This means that in the limit of vanishing coupling any spontaneous supersymmetry breaking originates in the O'Raifeartaigh model while the fermionic sector preserves supersymmetry.

\subsection{Expectation Values for Nonzero Interaction}
To calculate the expectation values for the component fields all auxiliary fields have to be eliminated from the superpotential. For the specific choice of structure constants outlined in Eqs. (\ref{ORaifeartaighconstlambda}) to (\ref{ORaifeartaighconstg}) the superpotential is given by
\begin{align}
U
	&=  F_1 F^\dagger_1
	+ F_2 F^\dagger_2
	+ F_3 F^\dagger_3
	- \frac{1}{2} M \phi_1 \phi_2
	- \frac{1}{2} M \bar{\phi}_1 \bar{\phi}_2 \notag \\
	&\quad - g A_1 \phi_1 \phi_3
	- \frac{1}{2} g A_3 \phi_1 \phi_1
	- g A^\dagger_1 \bar{\phi}_1 \bar{\phi}_3 
	- \frac{1}{2} g A^\dagger_3 \bar{\phi}_1 \bar{\phi}_1 \notag \displaybreak[3]\\
	&\quad + 2 \tilde{\lambda} \tilde{\lambda}
	+ \frac{1}{2} \tilde{\omega} \tilde{\omega}
	+ \left( \frac{m}{4} + \frac{\xi}{2} A_3 \right) \mathrm{Tr}{\left( \tilde{S}^T \tilde{S} \right)} \notag \displaybreak[3] \\
	&\quad + \left( \frac{m}{4} + \frac{\xi}{2} A_3^\dagger \right) \mathrm{Tr}{\left( \tilde{R}^T \tilde{R} \right)}
	+ 2 \tilde{\bar{\lambda}} \tilde{\bar{\lambda}}
	+ \frac{1}{2} \tilde{\bar{\omega}} \tilde{\bar{\omega}} \notag \displaybreak[3] \\
	&\quad + \left( \frac{m}{4} + \frac{\xi}{2} A_3^\dagger \right) \mathrm{Tr}{\left( \tilde{\bar{S}}^T \tilde{\bar{S}} \right)}
	+ \left( \frac{m}{4} + \frac{\xi}{2} A_3 \right) \mathrm{Tr}{\left( \tilde{\bar{R}}^T \tilde{\bar{R}} \right)} \notag \\
	&\quad + \xi \phi_3 \tilde{S} \chi
	- \xi \bar{\phi}_3 \tilde{\bar{S}} \bar{\chi}
	+ \xi \bar{\phi}_3 \tilde{R} \psi
	- \xi \phi_3 \tilde{\bar{R}} \bar{\psi} \, .
\end{align}
Inserting the equations of motion for the auxiliary fields $F_a$, $\tilde{\lambda}$, and $\tilde{\omega}$, as well as their hermitian conjugates leads to the on-shell superpotential
\begin{widetext}
\begin{align}
U
	&= M^2 A_2 A^\dagger_2
	+ 2 M g A_1 A^\dagger_2 A_3
	+ 2 M g A^\dagger_1 A_2 A^\dagger_3
	+ 4 g^2 A_1 A^\dagger_1 A_3 A^\dagger_3
	+ M^2 A_1 A^\dagger_1 
	+ \Lambda^2
	+ \Lambda g A_1 A_1
	- \xi \Lambda \left( \chi \chi + \bar{\psi} \bar{\psi} \right) \notag \\
	&\quad + \Lambda g A^\dagger_1 A^\dagger_1
	+ g^2 A_1 A_1 A^\dagger_1 A^\dagger_1
	- \xi g A^\dagger_1 A^\dagger_1 \left( \chi \chi + \bar{\psi} \bar{\psi} \right)
	- \xi \Lambda \left( \bar{\chi} \bar{\chi} + \psi \psi \right) 
	- \xi g A_1 A_1 \left( \bar{\chi} \bar{\chi} + \psi \psi \right) \notag \\
	&\quad + \xi^2 \left( \bar{\chi} \bar{\chi} + \psi \psi \right) \left( \chi \chi + \bar{\psi} \bar{\psi} \right)
	- \frac{1}{2} M \phi_1 \phi_2
	- \frac{1}{2} M \bar{\phi}_1 \bar{\phi}_2  
	- g A_1 \phi_1 \phi_3
	- \frac{1}{2} g A_3 \phi_1 \phi_1
	- g A^\dagger_1 \bar{\phi}_1 \bar{\phi}_3 
	- \frac{1}{2} g A^\dagger_3 \bar{\phi}_1 \bar{\phi}_1 \notag \\
	&\quad + 2 \left( \frac{m}{2} + \xi A_3 \right) \left( \frac{m}{2} + \xi A_3^\dagger \right) \chi \psi
	+ \left( \frac{m}{4} + \frac{\xi}{2} A_3 \right) \mathrm{Tr}{\left( \tilde{S}^T \tilde{S} \right)}
	+ \left( \frac{m}{4} + \frac{\xi}{2} A_3^\dagger \right) \mathrm{Tr}{\left( \tilde{R}^T \tilde{R} \right)} \notag \\
	&\quad + 2 \left( \frac{m}{2} + \xi A_3 \right) \left( \frac{m}{2} + \xi A_3^\dagger \right) \bar{\chi} \bar{\psi}
	+ \left( \frac{m}{4} + \frac{\xi}{2} A_3^\dagger \right) \mathrm{Tr}{\left( \tilde{\bar{S}}^T \tilde{\bar{S}} \right)}
	+ \left( \frac{m}{4} + \frac{\xi}{2} A_3 \right) \mathrm{Tr}{\left( \tilde{\bar{R}}^T \tilde{\bar{R}} \right)}
	+ \xi \phi_3 \tilde{S} \chi
	- \xi \bar{\phi}_3 \tilde{\bar{S}} \bar{\chi} \notag \\
	&\quad + \xi \bar{\phi}_3 \tilde{R} \psi
	- \xi \phi_3 \tilde{\bar{R}} \bar{\psi} \, .
\end{align}
\end{widetext}

A review of the equations of motion for $\tilde{\lambda}$ and $\tilde{\omega}$ reveals two possible solutions that lead to vanishing expectation values. First, there is the trivial solution with $\left\langle \chi_\alpha \right\rangle = \left\langle \psi_\alpha \right\rangle = 0$ which results in the same equations of motion for $F_a$ as the O'Raifeartaigh model. This leads to a nonvanishing expectation value for either $F_2$ or $F_3$. Therefore, supersymmetry is spontaneously broken. It can be shown that this scenario also has the same minimum of the superpotential as the O'Raifeartaigh model. This minimum is a local minimum as the superpotential acquires the finite positive value $\lambda^2$. 
Second, if the expectation value of $A_3 $ is chosen such that $\left\langle A_3 \right\rangle = - \frac{m}{2 \xi}$ the expectation values for the auxiliary fields $\tilde{\lambda}$ and $\tilde{\omega}$ vanish identically without making any assumptions on the component fields $\psi$ and $\chi$. Therefore, one or both of them can acquire a nonvanishing expectation value such that the expectation value for $F_3$ vanishes and supersymmetry is preserved. In this case the superpotential vanishes identically and thus represents a global minimum. It will be shown later on that restoring supersymmetry using nonvanishing expectation values of spinor products comes at the cost of breaking Lorentz invariance.

As the first case does not yield any new results the following discussion will be restricted to the second scenario. Starting from the equations of motion it can be shown that the component fields $A_a$ have the expectation values 
\begin{align}
\left\langle A_1 \right\rangle
	&= \left\langle A_2 \right\rangle
	= 0 \, , \,
	\left\langle A_3 \right\rangle
	= - \frac{m}{2 \xi} \, . 
\end{align}
Furthermore, the equations of motion imply a relation for the spinor fields $\chi$ and $\psi$
\begin{align}
\left\langle \bar{\chi} \bar{\chi} + \psi \psi \right\rangle
	&= \left\langle \chi \chi + \bar{\psi} \bar{\psi} \right\rangle
	= \frac{\Lambda}{\xi} \, .
\end{align}
These results then imply that the expectation values for the spinor fields $\phi_a$ vanish identically
\begin{align}
\left\langle \phi_1 \right\rangle
	&= \left\langle \phi_2 \right\rangle 
	= \left\langle \phi_3 \right\rangle 
	= 0 \, .
\end{align}
This does not come as a surprise as the expectation values for $\phi_a$ vanish in the O'Raifeartaigh model as well. In addition the component fields must satisfy the relations
\begin{align}
0
	&= \mathrm{Tr}{\left( \tilde{S}^T \tilde{S} + \tilde{\bar{R}}^T \tilde{\bar{R}} \right)} \, , \\
0
	&= \mathrm{Tr}{\left( \tilde{R}^T \tilde{R} + \tilde{\bar{S}}^T \tilde{\bar{S}} \right)} \, , \\
0
	&= \tilde{S}_{\alpha \beta} \chi^\beta + \tilde{\bar{R}}_{\alpha \dot{\beta}} \bar{\psi}^{\dot{\beta}} \, ,\\
0
	&= \tilde{\bar{S}}_{\dot{\alpha} \dot{\beta}} \bar{\chi}^{\dot{\beta}} + \tilde{R}_{\dot{\alpha} \beta} \psi^\beta \, .
\end{align}
This reveals that the second rank spinor fields $\tilde{R}$ and $\tilde{S}$ are not restricted and may acquire a nonzero expectation value
\begin{align}
\left\langle \tilde{R}_{\alpha \dot{\beta}} \right\rangle
	&= c_{R \alpha \dot{\beta}} \, , \, 
	\left\langle \tilde{S}_{\alpha \beta} \right\rangle
	= c_{S \alpha \beta} \, ,
\end{align}
where $c_{R \alpha \dot{\beta}}$ and $c_{S \alpha \beta}$ are constant second rank spinor fields.

\subsection{The Mass Terms}
The mass matrix is defined as the quadratic terms of the Lagrangian in the component fields after expanding them around their expectation values. In the previous section it was shown that $A_3$, $\chi$, $\psi$, $\tilde{S}$, and $\tilde{R}$ can acquire nonzero expectation values. Therefore, each of these fields has to be expanded. To distinguish the excitations from the component fields the bosonic fields are denoted by the corresponding small case italic letters, e.g. $A_1 \to a_1$, while the notation for fermionic excitations is extended by a hat, e.g. $\chi \to \hat{\chi}$. The expectation values, as long as they are not replaced by otherwise specified constants, are denoted by a subscript $0$. Therefore, the relevant terms of the superpotential are
\begin{widetext} 
\begin{align}
U_{\mathcal{O}^2}
	&= M^2 a_2 a^\dagger_2
	- \frac{m M g}{\xi} a_1 a^\dagger_2
	- \frac{m M g}{\xi} a^\dagger_1 a_2
	+ \frac{m^2 g^2}{\xi^2} a_1 a^\dagger_1
	+ M^2 a_1 a^\dagger_1
	+ \Lambda g a_1 a_1
	- \xi \Lambda \left( \hat{\chi} \hat{\chi} + \hat{\bar{\psi}} \hat{\bar{\psi}} \right)
	+ \Lambda g a^\dagger_1 a^\dagger_1 \notag \\
	&\quad - \xi g a^\dagger_1 a^\dagger_1 \left( \chi_0 \chi_0 + \bar{\psi}_0 \bar{\psi}_0 \right) 
	- \xi \Lambda \left( \hat{\bar{\chi}} \hat{\bar{\chi}} + \hat{\psi} \hat{\psi} \right)
	- \xi g a_1 a_1 \left( \bar{\chi}_0 \bar{\chi}_0 + \psi_0 \psi_0 \right)
	+ \xi^2 \bar{\chi}_0 \bar{\chi}_0 \hat{\chi} \hat{\chi}
	+ \xi^2 \bar{\chi}_0 \bar{\chi}_0 \hat{\bar{\psi}} \hat{\bar{\psi}}
	+ \xi^2 \psi_0 \psi_0 \hat{\chi} \hat{\chi} \notag \\
	&\quad + \xi^2 \psi_0 \psi_0 \hat{\bar{\psi}} \hat{\bar{\psi}} 
	+ 4 \xi^2 \bar{\chi}_0 \hat{\bar{\chi}} \chi_0 \hat{\chi}
	+ 4 \xi^2 \bar{\chi}_0 \hat{\bar{\chi}} \bar{\psi}_0 \hat{\bar{\psi}}
	+ 4 \xi^2 \psi_0 \hat{\psi} \chi_0 \hat{\chi}
	+ 4 \xi^2 \psi_0 \hat{\psi} \bar{\psi}_0 \hat{\bar{\psi}}
	+ \xi^2 \hat{\bar{\chi}} \hat{\bar{\chi}} \chi_0 \chi_0 
	+ \xi^2 \hat{\bar{\chi}} \hat{\bar{\chi}} \bar{\psi}_0 \bar{\psi}_0 
	+ \xi^2 \hat{\psi} \hat{\psi} \chi_0 \chi_0 \notag \\
	&\quad + \xi^2 \hat{\psi} \hat{\psi} \bar{\psi}_0 \bar{\psi}_0
	- \frac{1}{2} M \hat{\phi}_1 \hat{\phi}_2
	- \frac{1}{2} M \hat{\bar{\phi}}_1 \hat{\bar{\phi}}_2  
	+ \frac{m g}{4 \xi} \hat{\phi}_1 \hat{\phi}_1
	+ \frac{m g}{4 \xi} \hat{\bar{\phi}}_1 \hat{\bar{\phi}}_1
	+ 2 \xi^2 a_3 a_3^\dagger \chi_0 \psi_0
	+ \frac{\xi}{2} a_3 \mathrm{Tr}{\left( \tilde{S}_0^T \hat{\tilde{S}} + \hat{\tilde{S}}^T \tilde{S}_0 \right)} \notag \\
	&\quad + \frac{\xi}{2} a_3^\dagger \mathrm{Tr}{\left( \tilde{R}_0^T \hat{\tilde{R}} + \hat{\tilde{R}}^T \tilde{R}_0 \right)} 
	+ 2 \xi^2 a_3 a_3^\dagger \bar{\chi}_0 \bar{\psi}_0
	+ \frac{\xi}{2} a_3^\dagger \mathrm{Tr}{\left( \tilde{\bar{S}}_0^T \hat{\tilde{\bar{S}}} + \hat{\tilde{\bar{S}}}^T \tilde{\bar{S}}_0 \right)}
	+ \frac{\xi}{2} a_3 \mathrm{Tr}{\left( \tilde{\bar{R}}_0^T \hat{\tilde{\bar{R}}} + \hat{\tilde{\bar{R}}}^T \tilde{\bar{R}}_0 \right)}
	+ \xi \hat{\phi}_3 \tilde{S}_0 \hat{\chi} \notag \\
	&\quad + \xi \hat{\phi}_3 \hat{\tilde{S}} \chi_0
	- \xi \hat{\bar{\phi}}_3 \tilde{\bar{S}}_0 \hat{\bar{\chi}}
	- \xi \hat{\bar{\phi}}_3 \hat{\tilde{\bar{S}}} \bar{\chi}_0
	+ \xi \hat{\bar{\phi}}_3 \tilde{R}_0 \hat{\psi}
	+ \xi \hat{\bar{\phi}}_3 \hat{\tilde{R}} \psi_0
	- \xi \hat{\phi}_3 \tilde{\bar{R}}_0 \hat{\bar{\psi}}
	- \xi \hat{\phi}_3 \hat{\tilde{\bar{R}}} \bar{\psi}_0 \, .
\end{align}
\end{widetext}
Using the equation of motion for $F_3$ from Eq. (\ref{CouplingeqmF3}) simplifies the second order terms of the superpotential significantly
\begin{widetext}
\begin{align}
U_{\mathcal{O}^2}
	&= M^2 a_2 a^\dagger_2
	- \frac{m M g}{\xi} a_1 a^\dagger_2
	- \frac{m M g}{\xi} a^\dagger_1 a_2
	+ \frac{m^2 g^2}{\xi^2} a_1 a^\dagger_1
	+ M^2 a_1 a^\dagger_1 
	+ 4 \xi^2 \bar{\chi}_0 \hat{\bar{\chi}} \chi_0 \hat{\chi}
	+ 4 \xi^2 \bar{\chi}_0 \hat{\bar{\chi}} \bar{\psi}_0 \hat{\bar{\psi}}
	+ 4 \xi^2 \psi_0 \hat{\psi} \chi_0 \hat{\chi} \notag \\
	&\quad + 4 \xi^2 \psi_0 \hat{\psi} \bar{\psi}_0 \hat{\bar{\psi}} 
	- \frac{1}{2} M \hat{\phi}_1 \hat{\phi}_2
	- \frac{1}{2} M \hat{\bar{\phi}}_1 \hat{\bar{\phi}}_2  
	+ \frac{m g}{4 \xi} \hat{\phi}_1 \hat{\phi}_1 
	+ \frac{m g}{4 \xi} \hat{\bar{\phi}}_1 \hat{\bar{\phi}}_1
	+ 2 \xi^2 a_3 a_3^\dagger \chi_0 \psi_0
	+ \frac{\xi}{2} a_3 \mathrm{Tr}{\left( \tilde{S}_0^T \hat{\tilde{S}} + \hat{\tilde{S}}^T \tilde{S}_0 \right)} \notag \displaybreak[3] \\
	&\quad + \frac{\xi}{2} a_3^\dagger \mathrm{Tr}{\left( \tilde{R}_0^T \hat{\tilde{R}} + \hat{\tilde{R}}^T \tilde{R}_0 \right)}
	+ 2 \xi^2 a_3 a_3^\dagger \bar{\chi}_0 \bar{\psi}_0
	+ \frac{\xi}{2} a_3^\dagger \mathrm{Tr}{\left( \tilde{\bar{S}}_0^T \hat{\tilde{\bar{S}}} + \hat{\tilde{\bar{S}}}^T \tilde{\bar{S}}_0 \right)}
	+ \frac{\xi}{2} a_3 \mathrm{Tr}{\left( \tilde{\bar{R}}_0^T \hat{\tilde{\bar{R}}} + \hat{\tilde{\bar{R}}}^T \tilde{\bar{R}}_0 \right)}
	+ \xi \hat{\phi}_3 \tilde{S}_0 \hat{\chi} \notag \\ 
	&\quad + \xi \hat{\phi}_3 \hat{\tilde{S}} \chi_0
	- \xi \hat{\bar{\phi}}_3 \tilde{\bar{S}}_0 \hat{\bar{\chi}}
	- \xi \hat{\bar{\phi}}_3 \hat{\tilde{\bar{S}}} \bar{\chi}_0
	+ \xi \hat{\bar{\phi}}_3 \tilde{R}_0 \hat{\psi}
	+ \xi \hat{\bar{\phi}}_3 \hat{\tilde{R}} \psi_0
	- \xi \hat{\phi}_3 \tilde{\bar{R}}_0 \hat{\bar{\psi}}
	- \xi \hat{\phi}_3 \hat{\tilde{\bar{R}}} \bar{\psi}_0 \, .
\end{align}
\end{widetext}
At the same time it reveals an unpleasant multiplet structure as it turns out that there are three multiplets. Two of them are doublets, $a_1$ and $a_2$ as well as $\hat{\phi}_1$ and $\hat{\phi}_2$, containing fields with the same properties. The remaining fields -- $a_3$, $\hat{\chi}$, $\hat{\psi}$, $\hat{\phi}_3$, $\tilde{S}$ and $\tilde{R}$ -- form a sextuplet. This multiplet suffers of its large size as well as the mix of component fields it contains -- scalar fields, spinor fields and second rank spinor fields -- which makes it nearly impossible to reconcile.

A solution to the structure and size problem of this multiplet presents itself if it is recalled that the second rank spinor fields $\tilde{R}$ and $\tilde{S}$ were restricted by a relation involving the trace. This implies a constant expectation value which may or may not be zero. If it is assumed that the expectation values of both second rank spinor fields vanish identically the multiplet structure problem is resolved. The superpotential in second order of the component fields is reduced to
\begin{align}
U_{\mathcal{O}^2}
	&= M^2 a_2 a^\dagger_2
	- \frac{m M g}{\xi} a_1 a^\dagger_2
	- \frac{m M g}{\xi} a^\dagger_1 a_2
	+ \frac{m^2 g^2}{\xi^2} a_1 a^\dagger_1 \notag \\
	&\quad + M^2 a_1 a^\dagger_1
	+ 4 \xi^2 \bar{\chi}_0 \hat{\bar{\chi}} \chi_0 \hat{\chi} 
	+ 4 \xi^2 \bar{\chi}_0 \hat{\bar{\chi}} \bar{\psi}_0 \hat{\bar{\psi}} \notag \displaybreak[3] \\
	&\quad + 4 \xi^2 \psi_0 \hat{\psi} \chi_0 \hat{\chi}
	+ 4 \xi^2 \psi_0 \hat{\psi} \bar{\psi}_0 \hat{\bar{\psi}}
	- \frac{1}{2} M \hat{\phi}_1 \hat{\phi}_2 \notag \displaybreak[3] \\
	&\quad - \frac{1}{2} M \hat{\bar{\phi}}_1 \hat{\bar{\phi}}_2  
	+\frac{m g}{4 \xi} \hat{\phi}_1 \hat{\phi}_1 
	+ \frac{m g}{4 \xi} \hat{\bar{\phi}}_1 \hat{\bar{\phi}}_1
	+ 2 \xi^2 a_3 a_3^\dagger \chi_0 \psi_0 \notag \displaybreak[3] \\
	&\quad + 2 \xi^2 a_3 a_3^\dagger \bar{\chi}_0 \bar{\psi}_0
	+ \xi \hat{\phi}_3 \hat{\tilde{S}} \chi_0
	- \xi \hat{\bar{\phi}}_3 \hat{\tilde{\bar{S}}} \bar{\chi}_0
	+ \xi \hat{\bar{\phi}}_3 \hat{\tilde{R}} \psi_0 \notag \\
	&\quad - \xi \hat{\phi}_3 \hat{\tilde{\bar{R}}} \bar{\psi}_0 \, .
\end{align}
These terms result in a simpler multiplet structure which groups the component fields into four multiplets -- two doublets and two triplets. Three of them -- the doublets containing $\hat{\phi}_1$ and $\hat{\phi}_2$, $\hat{\chi}$ and $\hat{\psi}$, as well as the triplet containing $a_1$, $a_2$, and $a_3$ -- are easily explained and are solely made up of either scalar or spinor fields. The remaining triplet is slightly more involved as it groups a spinor field together with two second rank spinor fields. However, this mismatch can be resolved by the introduction of the fields
\begin{align}
\hat{\zeta}_{1 \alpha}
	&= - \hat{\tilde{S}}_{\alpha \beta} \chi_0^\beta \, , \\
\hat{\zeta}_{2 \alpha}
	&= \hat{\tilde{\bar{R}}}_{\alpha \dot{\beta}} \bar{\psi}_0^{\dot{\beta}} \, .
\end{align}
The terms involving $\phi_3$, $\tilde{R}$ and $\tilde{S}$ can then be expressed as
\begin{align}
U_{\phi_3 ,\tilde{S} , \tilde{R}}
	&= \xi \hat{\phi}_3 \hat{\zeta}_{1 \alpha}
	+ \xi \hat{\bar{\phi}}_3 \hat{\bar{\zeta}}_{1 \alpha}
	- \xi \hat{\bar{\phi}}_3 \hat{\bar{\zeta}}_{2 \alpha}
	- \xi \hat{\phi}_3 \hat{\zeta}_{2 \alpha} \, .
\end{align}
The previously mixed multiplet is transformed into a multiplet that contains the three spinor fields $\hat{\phi}_3$, $\hat{\zeta}_1$, and $\hat{\zeta}_2$. It has to be emphasized that this redefinition of fields is only successful after setting the expectation values for the second rank spinor fields to zero.

To determine the mass matrices it is still necessary to separate the complex scalar fields into their real components
\begin{align}
a
	&= \hat{a} + i \hat{b} \, .
\end{align}
At this point all multiplets but the one containing $\hat{\psi}$ and $\hat{\chi}$ are in a simple form. The remaining multiplet involves the products of four spinor fields including terms of the form $\chi_0 \chi_0 \chi \chi$ as well as cross terms $\chi_0 \chi \chi_0 \chi$.

Now it has to be recalled that at least one of the spinor fields $\chi$ and $\psi$ acquires a nonvanishing expectation value which satisfies $\Lambda / \xi = \chi \chi + \bar{\psi} \bar{\psi}$. Without loss of generality the spinor fields can be chosen such that $\chi$ acquires a finite expectation value while the expectation value for $\psi$ vanishes. Therefore, all but one term involving the product of four spinor fields vanish identically. Furthermore, the spinor field $\hat{\zeta}_2$ which is by definition proportional to the constant spinor field $\psi_0$ vanishes identically and the superpotential is simplified to
\begin{align}
U_{\mathcal{O}^2}
	&= \left( M^2 + \frac{m^2 g^2}{\xi^2} \right) \hat{a}_1 \hat{a}_1 
	+ M^2 \hat{a}_2\hat{a}_2
	- \frac{2 m M g}{\xi} \hat{a}_1 \hat{a}_2 \notag \\
	&\quad + \left( M^2 + \frac{m^2 g^2}{\xi^2} \right) \hat{b}_1 \hat{b}_1
	+ M^2 \hat{b}_2 \hat{b}_2 
	- \frac{2 m M g}{\xi} \hat{b}_1 \hat{b}_2 \notag \displaybreak[3] \\
	&\quad + 4 \xi^2 \bar{\chi}_0 \hat{\bar{\chi}} \chi_0 \hat{\chi}
	- \frac{1}{2} M \hat{\phi}_1 \hat{\phi}_2
	- \frac{1}{2} M \hat{\bar{\phi}}_1 \hat{\bar{\phi}}_2 \notag \\
	&\quad +\frac{m g}{4 \xi} \hat{\phi}_1 \hat{\phi}_1
	+ \frac{m g}{4 \xi} \hat{\bar{\phi}}_1 \hat{\bar{\phi}}_1
	+ \xi \hat{\phi}_3 \hat{\zeta}_{1 \alpha}
	+ \xi \hat{\bar{\phi}}_3 \hat{\bar{\zeta}}_{1 \alpha} \, .
\label{UCouplingfinal}
\end{align}
The bosonic component fields $\hat{a}_a$ and $\hat{b}_a$ can be grouped into two triplets of the form $a = \left( \hat{a}_1 , \hat{a}_2, \hat{a}_3 \right)$ and $b = \left( \hat{b}_1 , \hat{b}_2, \hat{b}_3 \right)$. The bosonic terms in Eq. (\ref{UCouplingfinal}) then correspond to the mass matrices
\begin{align}
M_a^2
	&= M_b^2
	= \begin{pmatrix}
	M^2 + \frac{m^2 g^2}{\xi^2} & - \frac{m M g}{\xi} & 0 \\
	- \frac{m M g}{\xi} & M^2 & 0 \\
	0 & 0 & 0
	\end{pmatrix} \, ,
\end{align}
which can be found in the reference literature, see, e.g. \cite{sohnius85}. The sole and important difference is that the corrections to the mass are no longer proportional to the scale parameter $\mu$ that sets the scale of the spontaneous supersymmetry breaking expectation value in the O'Raifeartaigh model. Instead the corrections to the mass term are proportional to the coupling strength $\xi$ between the O'Raifeartaigh model and the mass scale $m$ of the fermionic sector. This behavior was expected from the dimensional analysis in Section \ref{SSDimAnalysis} as the coupling terms via the $F$-component suggested a connection between the expectation value of the spontaneously broken superfield and the mass scale of the fermionic sector.

To formulate the fermionic mass matrix the two-spinors and their hermitian conjugates need to be grouped into four-spinors of the form
\begin{align}
\Phi'_i
	&= \begin{pmatrix} \phi_{i \alpha} \\ \bar{\phi}_i^{\dot{\alpha}} \end{pmatrix} \, ,
\end{align}
where $i = 1, 2$. This leads to a mass matrix for the fermionic doublet $\Phi' = \left( \Phi'_1 , \Phi'_2 \right)$ of
\begin{align}
M_{\Phi'}
	&= \begin{pmatrix}
	\frac{m g}{4 \xi} & - \frac{M}{4} \\
	- \frac{M}{4} & 0
	\end{pmatrix} \, .
\end{align}
This deviates from the results for the O'Raifeartaigh model that contains a spinor triplet $\Phi'' = \left( \Phi''_1 , \Phi''_2 , \Phi''_3 \right)$. The spinor field $\Phi'_3$ that is missing form the previously outlined doublet is massless in the O'Raifeartaigh model and forms a spinor triplet with the spinor fields $\hat{\zeta}_1$ and $\hat{\zeta}_2$ of the fermionic extension. It can be shown that the superpotential from Eq. (\ref{UCouplingfinal}) leads to the matrix 
\begin{align}
C_\zeta
	&= \begin{pmatrix}
	0 & \frac{\xi}{2} & 0 \\
	\frac{\xi}{2} & 0 & 0 \\
	0 & 0 & 0
	\end{pmatrix} \, .
\end{align}
This matrix has only off diagonal entries and thus cannot represent a mass matrix. A closer look at the mass dimensions of the spinor fields reveals that two of them, $\zeta_1$ and $\zeta_2$ have mass dimension $5/2$ while the remaining spinor field $\phi_3$ has mass dimension $3/2$. Therefore, the matrix represents a coupling matrix between the component fields of the O'Raifeartaigh model and the fermionic sector.

This nearly concludes the discussion of the terms in Eq. (\ref{UCouplingfinal}) and leaves only one last term, containing the product of four spinor fields, to explain
\begin{align}
U_\chi
	&= 4 \xi^2 \bar{\chi}_0 \hat{\bar{\chi}} \chi_0 \hat{\chi} \, .
\end{align}
Even though this term is to second order in the component fields, it is not a mass term. Instead, it is responsible for the breaking of Lorentz invariance as the contraction over the dotted and undotted spinor indices results in a term that is proportional to a vector-field while the spinor indices are absorbed into a $\sigma$-matrix. This results in a preferred direction which can conveniently be chosen in the time direction or in other words proportional to $\sigma^0$.

\section{\label{Sconclusion}Conclusions}
The primary objective of this article was to discuss the coupling of a supersymmetric model for fermionic fields with mass dimension one to the O'Raifeartaigh model.

Up to now no supersymmetric partners of SM particles were found. Therefore, it can be assumed that any realistic theory that is able to describe physics below the TeV scale must break supersymmetry either spontaneously or explicitly. This motivated utilizing the O'Raifeartaigh model to formulate a toy model. Subsequently, it has been shown that the only possible couplings of the fermionic sector to the O'Raifeartaigh model involving three fields contain one chiral superfield from the O'Raifeartaigh model as well as two superfields from the fermionic sector. Interestingly, contributions via the $F$-component are possible as well as various other terms that were neglected for the discussion. Moreover, these terms are similar to the mass terms that were used to construct the Lagrangian for the fermionic sector thus hinting at a possible connection between the vacuum expectation value of the spontaneously broken superfield and the mass scale of the fermionic sector.

As an example the coupling of the fermionic sector to the field with nonvanishing expectation value was discussed in detail. It is shown that the equations of motion for the auxiliary fields are very similar to those derived for the individual models. However, the $F$-term that corresponds to the field with nonvanishing expectation value acquires an additional contribution that is proportional to the coupling strength while the equations of motion for the fermionic sector acquire additional terms as well. A simple consistency check was performed by assuming a vanishing coupling strength. In this scenario it has been shown that the equations of motion of the coupled model reduce exactly to those of the individual models.

To find the expectation values for the component fields of the coupled model the superpotential had to be minimized. A brief look at the equations of motion for $\tilde{\lambda}$ and $\tilde{\omega}$ reveals that two distinct solutions exist. The first solution is the trivial solution $\left\langle \chi \right\rangle = \left\langle \psi \right\rangle = 0$ that leads to exactly the same equations of motion for $F_i$ as the O'Raifeartaigh model and, therefore, supersymmetry is spontaneously broken. The second solution where $\left\langle A_3 \right\rangle = -\frac{m}{2 \xi}$ is more intriguing. In this case, the equations of motion for $\tilde{\lambda}$ and $\tilde{\omega}$ are satisfied without making any assumptions on $\chi$ and $\psi$. If it is assumed that the two spinor fields have the finite expectation value such that $\left\langle \bar{\chi} \bar{\chi} + \psi \psi \right\rangle = \Lambda / \xi$ supersymmetry is restored. At the same time a nonvanishing expectation value for at least one of the spinor fields introduces a preferred direction and therefore breaks Lorentz invariance. Overall it is found that the scalar field $A_3$, the spinor fields $\chi$ and $\psi$, as well as the second-rank spinor fields $\tilde{R}$ and $\tilde{S}$ could acquire nonvanishing expectation values. 

To determine the mass matrices the component fields were expanded around their expectation values. It was found that a reasonable multiplet structure exists if and only if the expectation values for the second-rank spinor fields vanish identically. In this case the multiplet structure reduces to two fermionic doublets of which one is massless and two bosonic triplets. Furthermore, a fermionic triplets arises that combines spinor fields with different mass dimensions which makes it an interaction multiplet.

If the results for the coupled model are compared to those of the O'Raifeartaigh model a number of interesting differences have to be pointed out. The coupling to the fermionic sector restores supersymmetry at the cost of breaking Lorentz invariance. It is also found that the bosonic mass terms are now proportional to the coupling strength as well as the mass scale of the fermionic sector and no longer dependent on the arbitrary scale parameter of the O'Raifeartaigh model. The fermionic triplet of the O'Raifeartaigh model is replaced by a fermionic doublet that also depends on the coupling strength and the mass scale of the fermionic sector. Furthermore, there is an additional massless fermionic doublet. Finally, a fermionic triplet containing spinors with different mass dimension exists. It represents a coupling matrix that does not have an equivalent in the O'Raifeartaigh model.

At this point the motivation behind the specific choice of coupling between the fermionic sector and the O'Raifeartaigh model becomes clear. Up to now no superpartners were detected experimentally, therefore, supersymmetry must be broken at energies currently accessible to experiments. Furthermore, the coupling to the fermionic sector mimics a coupling to the Higgs field of the SM. Therefore, an extension of the presented formalism to an extension of the SM or MSSM should in general be possible in perfect analogy and potentially result in similar effects. Due to mass dimensional arguments the coupling of the fermionic sector to the Higgs field would dominate while all other couplings are suppressed. Therefore, the presented model for fermionic fields with mass dimension one provides a good candidate for supersymmetric dark matter. Provided the Higgs particle is detected at the LHC, potential deviations from the expected branching ratios of the Higgs particle could then, at least in principle, be used to predict the mass scale and coupling strength of the fermionic sector thus providing experimental constraints on the amount of supersymmetric dark matter.

These results show that the presented model for fermionic fields with mass dimension one is an interesting candidate for supersymmetric dark matter that could be accessible to experiments in the near future.

\begin{acknowledgments}
This work was supported by the Natural Science and Engineering Research Council, Canada. 
\end{acknowledgments}

\bibliography{KEW4PRD}

\end{document}